\documentclass[prl,twocolumn]{revtex4-1}
\usepackage{amsmath,amssymb,bm}

\newcommand{\ket}[1]{|{#1} \rangle }

\usepackage{amsmath}	
\usepackage{color}
\usepackage[pdftex]{graphicx}
\usepackage{hyperref}
\usepackage[normalem]{ulem}

\begin{document}
	\title{The $^{9}$Li(d,p) reaction, specific probe of $^{10}$Li, paradigm of parity--inverted nuclei around $N=6$ closed shell}
	\author{F. Barranco}
	\affiliation{Departamento de F\`isica Aplicada III,
Escuela Superior de Ingenieros, Universidad de Sevilla, Camino de los Descubrimientos, 	Sevilla, Spain}	
	\author{G. Potel}
	\affiliation{National Superconducting Cyclotron Laboratory, Michigan State University, East Lansing, Michigan 48824, USA}
	\author{E. Vigezzi}
\affiliation{INFN Sezione di  Milano,
Via Celoria 16, 
I-20133 Milano, Italy }	
	\author{R. A. Broglia}
	\affiliation{ The Niels Bohr Institute, University of Copenhagen, 
DK-2100 Copenhagen, Denmark }
\affiliation{Dipartimento di Fisica, Universit\`a degli Studi Milano,
Via Celoria 16, 
I-20133 Milano, Italy }

	\date{\today}
	\begin{abstract}
		We show, within the framework of renormalized nuclear field theory and of the induced reaction surrogate formalism, that the highly debated $^{10}$Li structure, 
	studied in a recent high statistics 	$^9$Li(d,p)$^{10}$Li  one--neutron transfer experiment,	
	is consistent 
	with,  or better, requires, the presence of a virtual $1/2^+$ state 
of similar single--particle strength than that of the $1/2^-$ resonance at 0.45$\pm0.03$ MeV,
provided one looks for $E_x \leq $ 0.2 MeV,  at a quite different angular region ($\theta_{cm}  > 40^{\circ}$) than 
that employed in the experiment (5.5$^{\circ}$ - 16.5$^{\circ}$),
in which the $s-$strength ($2^-,1^-$)
has a minimum, while the $p_{1/2}$ one ($1^+,2^+)$  displays a maximum.
Based on continuum spectroscopy self-energy techniques, we find that  the  physical mechanism responsible for  parity inversion in $^{10}_3$Li is the same as that at the basis of the similar phenomenon observed in  $^{11}_4$Be and as that needed in $^{11}$Li to have an important $s$--wave ground state component. Furthermore, it is also consistent with the (normal) sequence of the $1p_{1/2}$ and $2s_{1/2}$ levels in the $N=7$ isotones $^{12}_5$B and $^{13}_6$C. 
	\end{abstract}	
	\maketitle
	\textit{Introduction}. 


	A central issue in the study of atomic nuclei is that of the identification of the magic numbers. 
	Seven decades have elapsed since the seminal papers in which Mayer \cite{Mayer:48,Mayer:49,Mayer:49b} (see also \cite{Elsasser:33}) and Jensen \cite{Haxel:49} introduced the shell model of the atomic nucleus. Much work on the subject has been dedicated by the nuclear physics community ever since 
(\cite{Mayer:55,DeShalit:63,Bohr:69,Bertsch:72,Navratil:09,Vary:15} 
and references therein). In spite of this, the quest for the pillars of nuclear structure, namely the magic numbers, is far from completed, being very much an open question reserving surprises and challenges
\cite{Sorlin:08,Heide:11,Krucken:11,Hagen:12,Otsuka:16}. 

	\textit{Novel magic numbers: parity inversion}. The first two Meyer--Jensen magic numbers are 2 and 8 for both protons and neutrons. 
	
	Increasing the number of neutrons of a normal nucleus, Pauli principle forces them into states of higher momentum. When the core becomes neutron saturated, the nucleus expels most of the wave function of the last neutrons outside to form a halo which, because of its large size, can have a lower momentum, that is less kinetic energy of confinement. The system $^{11}_4$Be$_7$ ($(N-Z)/A\approx0.36$) constitutes a much studied example of one neutron halo nucleus (\cite{Winfield:01,Calci:16,Barranco:17} and refs. therein). In principle one could have expected that because the $1s_{1/2}$ and $1p_{3/2}$ are filled, the last neutron occupies a substate of the $1p_{1/2}$ orbital. 
However, the ground state of $^{11}$Be has spin and parity $1/2^+$, implying inversion in the sequence of the $1p_{1/2}$ and $2s_{1/2}$ orbitals. Because the $1/2^+$ (-0.50 MeV) and $1/2^-$ (-0.180 MeV) levels are very close to each other, and separated from the $3/2^-$ level by about 3 MeV the $N=8$ role of magic number has been taken over by $N=6$. In other words, $^{11}$Be can be viewed as a one--neutron system outside the closed shell, the reaction $^{10}$Be$(d,p)^{11}$Be being then the specific probe of such a system. It is furthermore of notice that closely associated with the parity inversion phenomenon, the dipole transition between the $1/2^+$ and $1/2^-$ states carries about one Weisskopf unit, being the strongest $E1$--transition between bound states of the whole mass table \cite{Kwan:14}.
A piece of information which can be used at profit in connection with the position of the $1/2^+$ and $1/2^-$ states in $^{11}$Li
(see below).

					\begin{figure}[t]
					{\includegraphics*[width=9cm,angle=0]{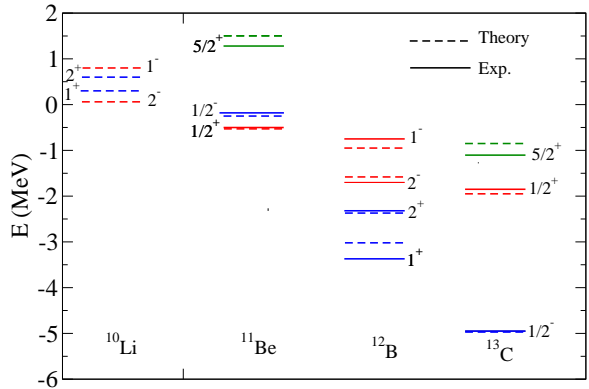}}
\caption{(color online)
The  experimental energies of the low-lying states in  the $N=7$ isotones  $^{11}$Be, $^{12}$B  and $^{13}$C  are shown by  solid lines.
The corresponding  theoretical  energies     are shown  by   dashed lines. Also shown are the predictions in $^{10}$Li.
States based on $1/2^-,1/2^+$ and $5/2^+$ neutron configurations are shown by blue, red and green   lines respectively.
}
\label{fig1}
\end{figure}


A substantial set of experimental data \cite{Wilcox:75,Young:94,Bohlen:97,Caggiano:99,Santi:03,Jeppesen:06,Smith:15,Sanetullaev:16,
Kryger:93,Zinser:95, Kobayashi:97,Thoennessen:99,Chartier:01,Simon:07,Aksyutina:13,Amelin:90,Gornov:98,Chernysev:15,Kanungo:15,
Tanihata:13,Smith:15,Sanetullaev:16,Fortune:18,Tanihata:08}
and  theoretical insight \cite{Thompson:94,Bertsch:98,Blanchon:07,Gomez:17, Barker:77,Poppelier:93,Kitagawa:93,
Descouvemont:97,Wurzer:96,Kato:93,Thompson:94,Garrido:02,Garrido:03,Potel:10,Casal:17,
Vinh:95,Nunes:96,Sagawa:93,Myo:08}
 exists on the unbound isotone of $_4^{11}$Be, namely $_3^{10}$Li which indicates parity inversion also in this case. 	
 This scenario is furthermore consistent  with --required by -- the bound, two neutron halo system $^{11}_3$Li$_8$ (\cite{Barranco:01} and 
 refs. therein). The presence of a low--lying dipole resonance with $\approx6-8$\% of the dipole Energy Weighted Sum Rule (EWSR), of  $\approx0.5$ MeV  width and centroid energy $\lesssim1$ MeV (\cite{Kanungo:15} and refs. therein), and thus carrying about one Weisskopf unit, implies the presence of a particle--hole dipole excitation with energy not much larger than 0.3-0.5 MeV. Furthermore, the value of the absolute differential two--neutron pickup cross section associated with the reaction $^1$H($^{11}$Li,$^9$Li(gs))$^3$H \cite{Tanihata:08} implies that the $\ket{s_{1/2}^2(0)}$ and $\ket{p_{1/2}^2(0)}$ configurations enter the $^{11}$Li ground state with about the same amplitude (0.45 $\ket{s_{1/2}^2(0)}$+ 0.55$\ket{p_{1/2}^2(0)}$ \cite{Potel:10}, the $1p_{3/2}(\pi)$ odd proton being considered as a frozen spectator, is not explicitly written).
The above requirement implies that the energies of the two configurations are not too different, likely within 0.5--0.6 MeV.  

The generally accepted picture  was recently set in doubt as a result of a one-nucleon transfer 
experiment \cite{Cavallaro:17}  which led to the  conclusion that  "... the level sequence in the $^{10}$Li system may 
not show the shell inversion  features observed in other N=7 isotopes such as $^{11}$Be".  The specificity
of the $^9$Li(d, p) reaction to provide insight into the level sequence of a nuclear system combined with the high statistics 
of the experiment, implied that the above conclusions constitute a serious question mark on the validity of the entire picture 
of shell evolution leading to the N=6 magic number and associated parity inversion in exotic halo nuclei 
at the neutron drip line, and of the crucial importance of finding the missing link. This is the challenge we take up in
the present paper.  And to do so we use theoretical tools in which structure and reactions and thus bound states 
and continuum dynamics become unified.

\textit{Scope and outcome}. In this letter we will show that the data of Cavallaro \textit{et al.} \cite{Cavallaro:17} are consistent with  the presence of $1/2^+$ strength at threshold of similar magnitude as that of the $1/2^-$ resonance observed at $0.45\pm0.03$ MeV. 
But to observe it one has to look into a rather different   angular region than that used in the experiment 
(5.5$^{\circ}$-16.5$^{\circ}$), involving angles $\theta_{cm}>40^{\circ}$ and likely centred at backward angles, said it differently 
implying a different region of momentum transfer which particularly privileges non-spin flip configurations. 
Thus, essentially the state $|(s_{1/2} \otimes p_{3/2}(\pi))_{2^-}>$.
Detailed predictions of the full angular distribution and of the absolute energy differential cross sections associated with the $2s_{1/2}$, $1p_{1/2}$, $1p_{3/2}$ and $1d_{5/2}$ virtual and resonant states are also given.  

The theoretical framework used provides, at the same time, a unified account of the experimental findings regarding 
	the $N=7$ isotopes $^{13}_6$C, $^{12}_5$B, $^{11}_4$Be and $^{10}_3$Li, and closes the issue concerning the
	missing $s$-strength at threshold in $^{10}$Li.

\textit{Transferability}.  Because the mechanism at the basis of the phenomenon of parity inversion observed in $^{11}$Be associated with   the dressing
of the $1p_{1/2}$ and $2s_{1/2}$ single-particle states, through the coupling of the quadrupole vibration of the core $^{10}$Be (self-energy processes),
namely the coupling of the $\ket{2s_{1/2}}$ state with the $\ket{1d_{5/2}\otimes2^+;1/2^+}$ configuration and of the $\ket{1p_{1/2}}$ with the $\ket{((p_{1/2},p_{3/2}^{-1})_{2^+}\otimes 2^+)_{0^+}, p_{1/2};1/2^-}$ configuration, leading to attraction and to repulsion respectively \cite{Barranco:17},
 is equally operative in $^{10}$Li than in $^{11}$Be, it is sensible to assume that the $^{11}$Be observations and theoretical results are transferable to $^{10}$Li. Similarly, because successive transfer is the main mechanism in the $^1$H($^{11}$Li,$^9$Li(gs))$^3$H process \cite{Potel:10}, the properties of the dressed single--particle states entering $^{11}$Li are also likely to be transferable to $^{10}$Li.
	


	\textit{Methods and results}. In the calculation of $^{11}_4$Be \cite{Barranco:17}, the four parameters characterizing the bare mean field $U(r)$ (to be used in connection with an effective mass $m_k(r)(m_k=0.7 m (0.91 m)$ for $r=0$ $ (r=\infty)$),were determined imposing the self--consistent condition that the dressed single--particle levels resulting from the coupling to the quadrupole vibration of the core $^{10}$Be reproduce the experimental energies, in particular those of the parity inverted $1/2^+$ and $1/2^-$ states. We have extended this approach to the normal sequence of  $N=7$ isotones $^{12}_5$B and $^{13}_6$C (see Suppl. Mat.), 
	obtaining again an accurate reproduction of the experimental spectra (see Fig. 1).

The effect of the neutron-proton interaction, leading to the observed doublet splitting in $^{12}$B,
is discussed  below. 

The parameters of the bare potential have then been extrapolated   to the case of $^{10}$Li. With this global potential and $k$--mass, together with the quadrupole vibration of $^{9}$Li ($\hbar\omega_2=3.37$ MeV, $\beta_2=0.72$), we have calculated the corresponding single--particle renormalization processes in $^{10}$Li. 
Diagonalizing the associated self--energy  $\Sigma_{ik}(E)$ $(\epsilon_i,\epsilon_k>\epsilon_F)$ 
the dressed $\widetilde{1/2^+},\widetilde{1/2^-}$
 neutron states were calculated
 \footnote{Within this context, it has been stated (see e.g. \cite{Kuo:97}) that exotic (halo) nuclei being much less bound  than nuclei 
 lying along the stability valley, offer a unique framework to study mean field properties without the complications 
 of polarization from valence particles (separability issue). 
  In particular one could argue  that valence neutrons of $^{210}$Pb,
 can exert a much larger polarisation of the core $^{208}$Pb, than those of $^{11}$Li regarding the $^9$Li core,
 let alone in the case of $^{209}$Pb as compared with the extreme case of $^{10}$Li.
 Inverting the issue one could e.g., posit that the dressing of the valence neutron in $^{10}$Li due to the quadrupole vibration of the core is 
 poorer than that of the valence neutron of $^{209}$Pb due to a similar mechanism. Now, the total zero point amplitude 
 associated with the lowest mode of  $^{208}$Pb is $\beta_2 \approx 0.06$ (see e.g. \cite{Bohr:75} Eqs. (6-52) and (6-390)) while that of 
 $^9$Li is $\beta_2 \approx 0.66$, namely one order of magnitude larger. Because the dressing (self-energy) of single-particle states is proportional to $\beta^2_2$, the assumption of separability, essentially not applicable in the case of $^{209}$Pb, can hardly be used in treating 
 $^{10}$Li or, for that sake, in treating any of the  $N=7$ isotopes considered in the present paper
 (see also \cite{Broglia:10}). }.
 
  The virtual state $\widetilde{1/2^+}$  can be expressed as,
		\begin{equation}
		\label{eq1}
	 \ket{\widetilde  {1/2}^+}  =\sqrt{0.98}\ket{s_{1/2}}   
	 +\sqrt{0.02}\ket{\left(d_{5/2}\otimes 2^+\right)_{1/2^+}}.
	 \end{equation}
	 The associated scattering length  is $\alpha= - lim _{k\to 0} tg(\delta_{1/2^+})/k = - 8$ fm,
	 corresponding to the energy   \cite{Landau:81,Friedrich:16} $\epsilon_{\widetilde{1/2}^+} = 
	\frac{\hbar^2\kappa^2}{2m} 	=  0.32 \; {\rm MeV} $,
	 where $ \kappa = 1/\alpha$.
The resonant  $\widetilde{1/2^-}$ state can be written as 
	\begin{align}\label{eq2}
	& \ket{\widetilde  {1/2}^-}   = \nonumber  \\  & \sqrt{0.94}\ket{p_{1/2}}  
 +\sqrt{0.07} \;  \;\ket{((p_{1/2},p_{3/2}^{-1})_{2^+}\otimes 2^+)_{0^+},p_{1/2};1/2^-}.
 \end{align}
 The peak and the width of the resonance are $\epsilon_{\widetilde{1/2}^-} = $ 0.50 MeV, and 
$\Gamma_{\widetilde{1/2^-}}  =  2\left(\frac{d\delta_{1/2^-}}{dE}|_{\epsilon_{\widetilde{1/2^-}}}\right)^{-1}= 0.35 \; {\rm MeV}	.$
	
	


		The parallel between these results and those shown in Eqs (1)--(3) of ref. \cite{Barranco:17} for $^{11}$Be, let alone with those displayed in Fig. 1 of ref. \cite{Barranco:01} and in Eqs (1)--(4) of ref. \cite{Potel:10} is apparent. 

The dressed  neutron  
	couples to the $1p_{3/2}^{-1}(\pi)$ proton hole  leading
	to the doublets ($1^-,2^-$) (${\widetilde{1/2}^+}  \otimes  p_{3/2}^{-1}(\pi)$) and ($1^+$, $2^+$) (${\widetilde{1/2}^-}  \otimes p_{3/2}^{-1}(\pi)$),
	both in $^{10}$Li and $^{12}$B. 
      In keeping with  a well established approach \cite{Deshalit:53,Talmi:60,Heide:90}, 
	we assume that the proton 
         interacts with the odd neutron through a spin-dependent contact interaction, 
	$V_{np} = F_0 \delta(\mathbf r_1-\mathbf r_2) \left(1-\alpha(1 -\pmb\sigma_1\cdot\pmb\sigma_2)\right)$.
	The two parameters  $F_0$ and $\alpha$ are  determined by fitting the experimental 
	 splittings in $^{12}$Be (see Fig. 1)  and optimising the agreement with the transfer cross sections measured  in $^{10}$Li as discussed below
	  (see also Suppl. Mat.).   

\begin{figure}[h]
				{\includegraphics*[width=4.2cm,angle=0]{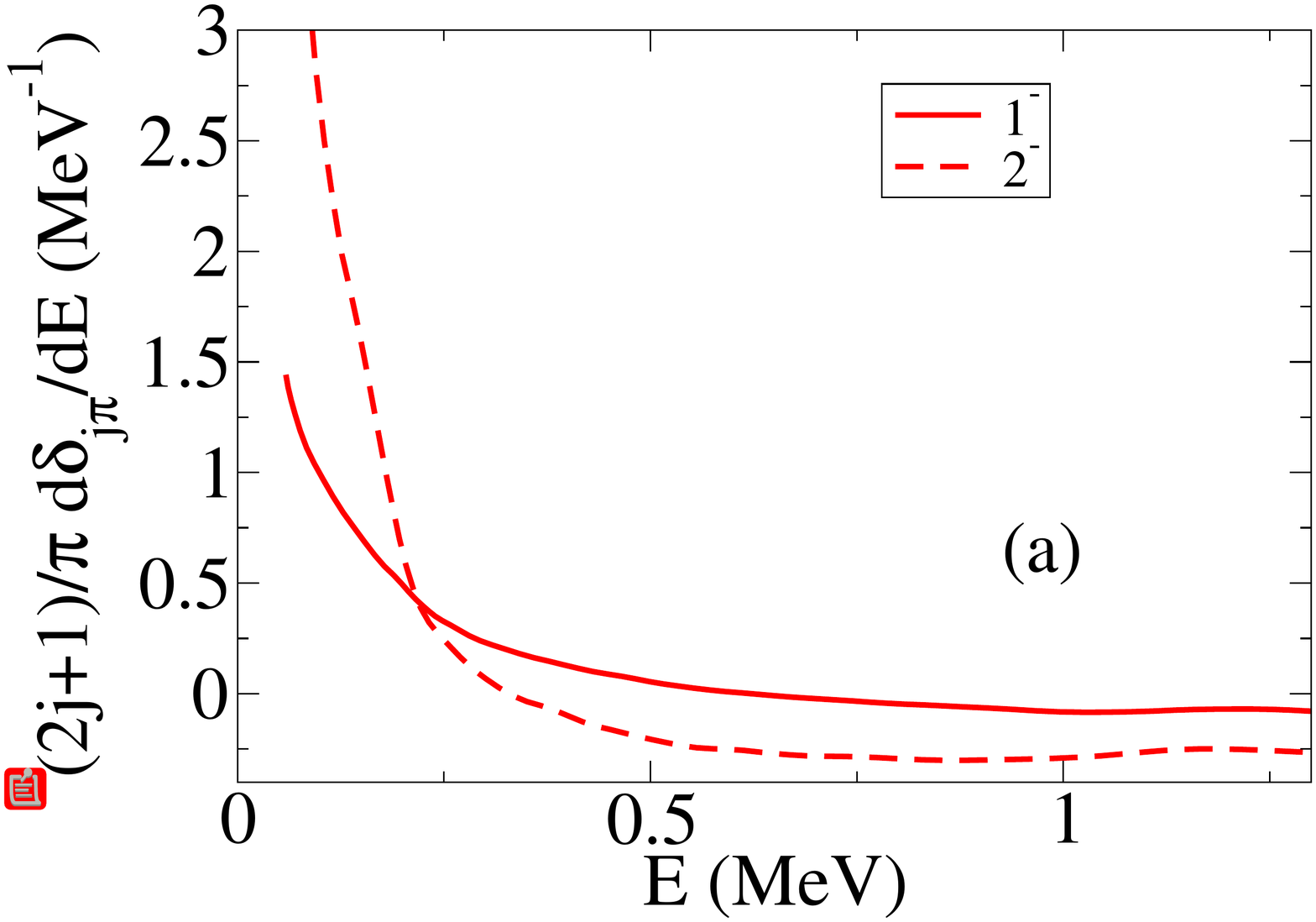}}
				{\includegraphics*[width=4.2cm,angle=0]{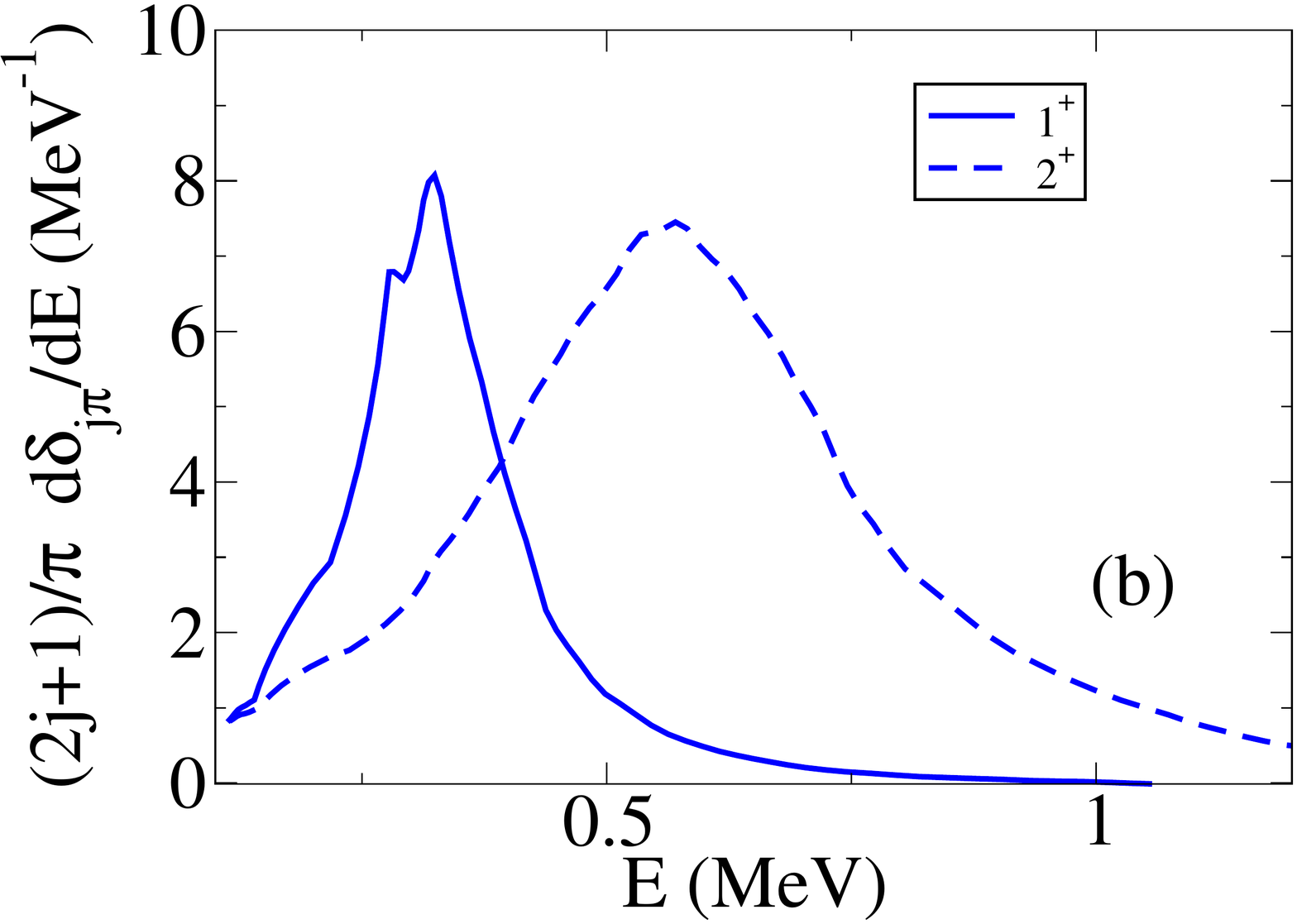}}				
\caption{ Spectral functions of the low-lying negative (a) and positive  (b) parity states in $^{10}$Li.}
\label{fig_spectral}
\end{figure}

The scattering lengths of the resulting  $2^-,1^-$ states  are equal to -19  fm  and -5 fm respectively ($\epsilon_{2^-} \approx$  0.05 MeV, $\epsilon_{1^-} \approx$  0.8 MeV).
For the positive parity doublet one finds instead  ($\epsilon_{1^+} \approx$  0.3 MeV, $\epsilon_{2^+} \approx$  0.6 MeV).
	  The  calculated spectral functions  of these four low-lying states are shown  in Fig. \ref{fig_spectral}. 
	  For states lying in the continuum  it 
	  is convenient to consider the change of the spectral function of the interacting system with respect to the free spectral function,
	  $\bar{\rho}_{j^{\pi}} (\omega)= -\frac{(2j+1)}{\pi} Im [G_{j^{\pi}}(\omega + i0^+)-   G_{0,j^{\pi}}(\omega + i0^+)] = \frac{(2j+1)}{\pi} \frac{d\delta_{j^{\pi}}}{d\omega}$
	  \cite{Beth:37,Shlomo:97,Mizuyama:12} (see also  \cite{Huang},p.226).
	  
Furthermore, theory  predicts  the existence  of a resonant $5/2^+$ state with centroid at $\approx3.5$ MeV 
which splits into four states $(\widetilde {5/2}^+ \otimes 1p_{3/2}(\pi))_{1^-,2^-,3^-,4^-}$ spanning the energy interval 2-5 MeV,
as well as  of a
 $3/2^-$  
 resonance
which splits into four states ($\widetilde{3/2}^- \otimes 1p_{3/2}(\pi)$)$_{0^+,1^+,2^+,3^+}$ with energies within the range 3--6 MeV. 
and of another one 
expected at $\approx$5.4 MeV, and based on the state $\ket{(p_{3/2}^{-1}\otimes 0^+_a)_{3/2^{-}}}$ where $\ket{0^+_a}=\ket{gs(^{11}\text{Li})}$ is the pair addition mode of the core $^{9}$Li, that is the ground state of $^{11}$Li. 
	
We estimate the coupling between the $3/2^{-}$ resonances to occur
	mainly through the ($p_{1/2}\otimes 2^+)_{3/2^-}$ configuration, and to be weak. Within this context we recall a similar situation, this time for bound states, concerning the two $3/2^+$ states found in connection with the study of the septuplet of states $\ket{h_{9/2}\otimes 3^-(^{208}\text{Pb});I}$ ($I=3/2^+,5/2^+,\dots 15/2^+$) of $^{209}$Bi, the second $3/2^+$ being connected with the 2$p-1h$ state $\ket{d_{3/2}^{-1}\otimes gs(^{210}\text{Po});3/2^+}$ (see \cite{Bortignon:77} and references therein). In this case the mixing between the two states is much larger due mainly   to the fact that the unperturbed energies of the two $3/2^+$ states are almost degenerate.

				\begin{figure}[h]
				{\includegraphics*[width=9cm,angle=0]{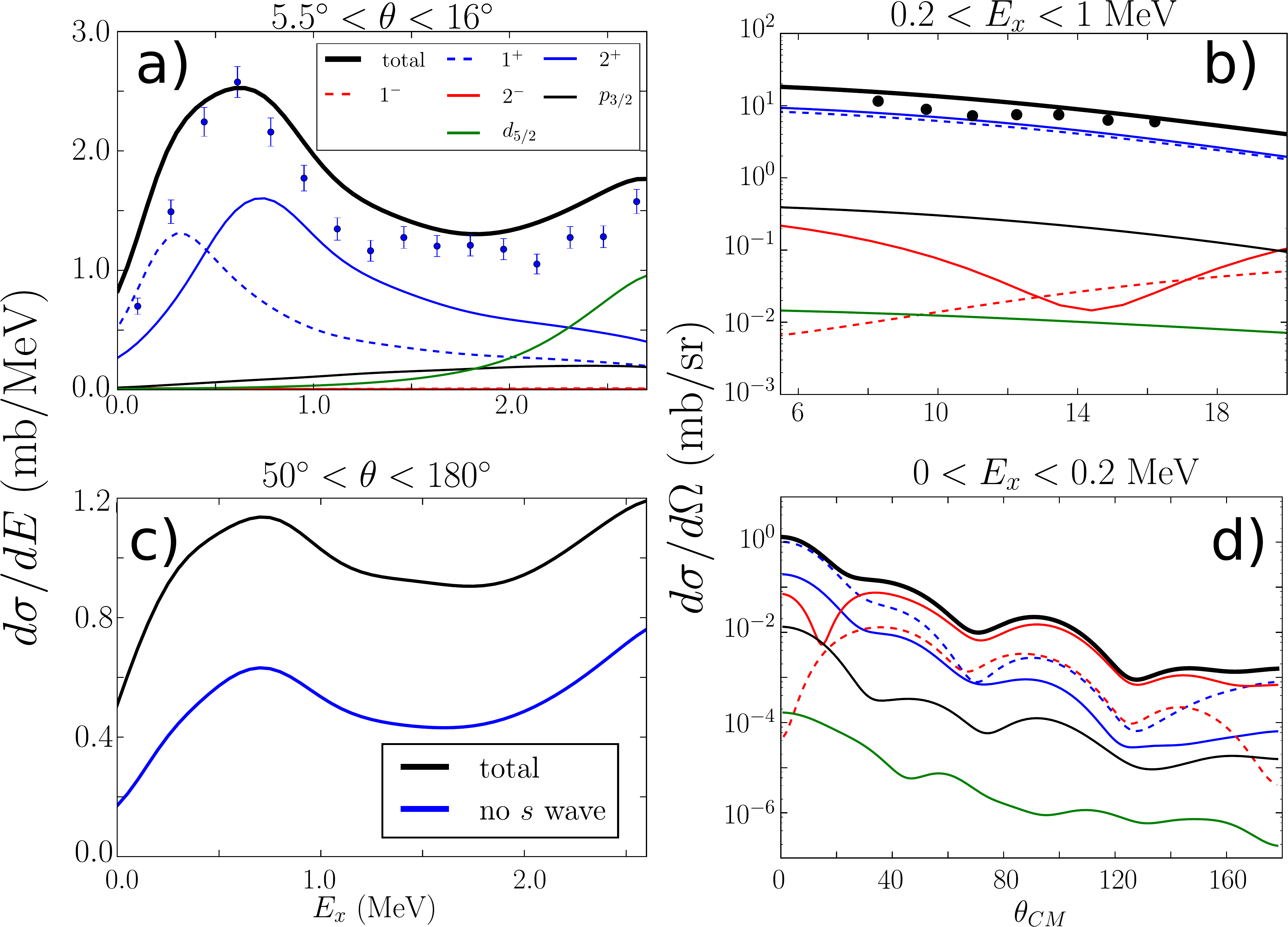}}
\caption{ (Color online) (a)  Theoretical prediction (continuous solid curve) of the $^{10}$Li strength function for the $d(^9$Li,$p)^{10}$Li reaction at 100 MeV incident energy and $\theta_{CM}=[5.5^\circ,16.5^\circ]$ in comparison with the experimental data (solid dots with errors) \cite{Cavallaro:17};  (b) Corresponding angular distributions associated with the states in the  energy interval 0.2--1 MeV in comparison with the experimental data; (c) Predicted strength function integrated in the  angular range $\theta_{CM}=[50^\circ,180^\circ]$, compared to the result obtained neglecting the contributions from the s-wave ($1^-$ and $2^-$ states); (d) Predicted angular distributions integrated in the energy interval  0--0.2 MeV. The contributions to the cross sections   associated with the  $1/2^+$,$1/2^-$, $3/2^-$ and $5/2^+$ states
are also shown in panels (a),(b) and (d).
}
\label{fig2}
\end{figure}

	\begin{figure}[h]
	{\includegraphics*[width=9cm,angle=0]{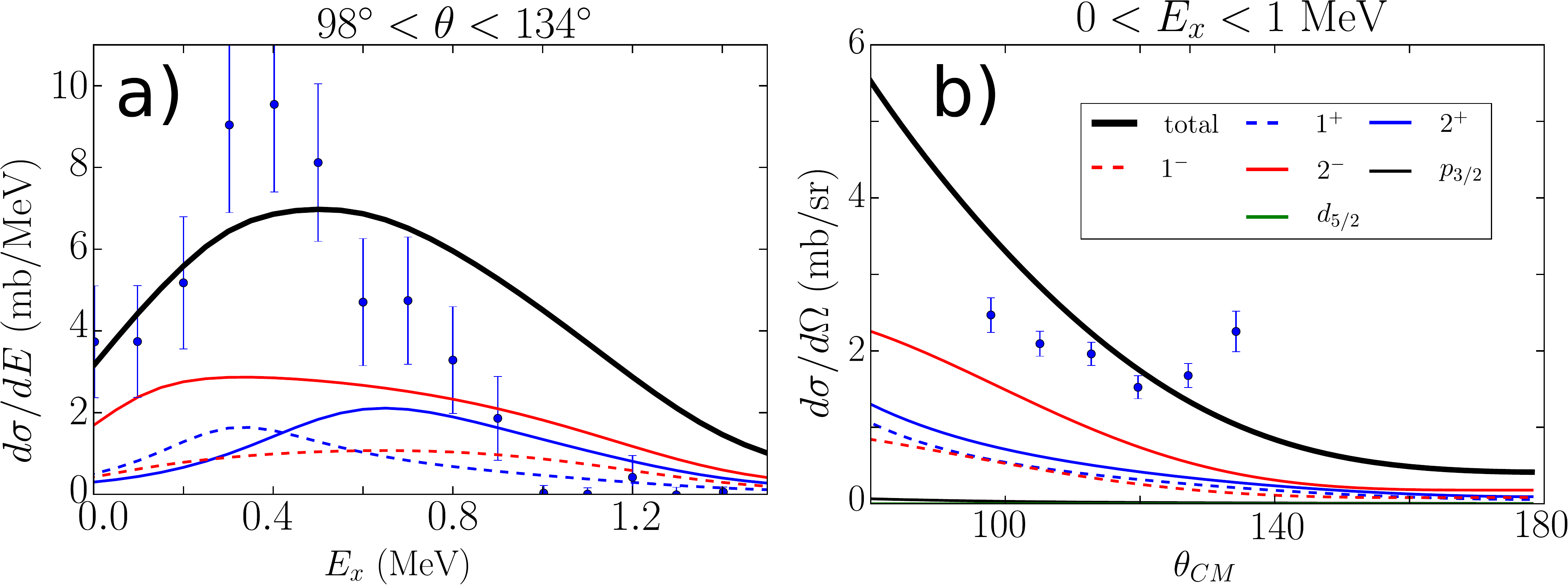}}
		\caption{(Color online) (a)  Theoretical prediction (continuous solid curve) of the $^{10}$Li strength function for the $d(^9$Li,$p)^{10}$Li reaction at 21.4 MeV incident energy and $\theta_{CM}=[98^\circ,134^\circ]$ in comparison with the experimental data.  (solid dots with errors) \cite{Jeppesen:06};  (b) Corresponding angular distributions associated with the states in the  energy interval 0--1 MeV in comparison with the experimental data	. It is of notice that while the experimental data displayed in (a) are reported in arbitrary units \cite{Jeppesen:06}  those shown in (b)
		are given in mb/sr. The theoretical predictions are in all cases in absolute values. }
\label{fig3}
	\end{figure}
			


Based on the non--local self--energy matrices $\Sigma(r,r';E)$, whose configuration space representation corresponds to $\Sigma_{ik}(E)$, in conjunction with the optical parameters of ref. \cite{Schmitt:13}, we have calculated the absolute double differential cross section $d^2\sigma/dEd\Omega$ within the framework of induced--surrogate reaction formalism (\cite{Potel:15b,Potel:17b} and refs. therein). The associated absolute single differential cross sections $(d\sigma/dE)_{5.5^\circ-16.5^\circ}$ and $(d\sigma/d\Omega)_{0.2-1{\text{ MeV}}}$  were obtained by integrating $d^2\sigma/dEd\Omega$ in the angular and energy ranges within which data was recorded  \cite{Cavallaro:17}.  In keeping with the experimental energy resolution, the theoretical results were folded with Lorentzian functions of FWHM of 170 keV.

	  As seen from Figs. \ref{fig2} (a) and (b), theory provides a quantitative account of the experimental findings, confirming the lack of any 
	  relevant contribution associated with  $s_{1/2}$ strength and thus the doubts regarding the presence of parity inversion in $^{10}$Li. 
	  Very similar results were obtained making use of the optical parameters employed in ref. \cite{Cavallaro:17}.

	The picture changes radically when looking at a quite different angular region, this time centered  around more backward 
	angles ($\theta \geq 40^{\circ}$), and thus corresponding to larger momentum transfer,
as testified by the  cross sections     $(d\sigma/dE)$ integrated in the
 angular range ${50^\circ-180^\circ}$  (Fig. \ref{fig2} (c)) as well as by the absolute differential cross section at angles $\theta_{CM}> 40^o$  (Fig. \ref{fig2} (d)),
	which unarguably demonstrate the presence of a virtual $s_{1/2}$ state, state absent (non observable) from $(d\sigma/dE)_{5.5^\circ-16.5^\circ}$.
	To make such a statement it is required to be able to predict absolute one-particle transfer cross sections within experimental errors. To fulfil such requirements 
	one has to able to calculate  continuum self-energy processes. That is, the dressing of a particle state through the coupling of the quadrupole vibration
	of the core $^9$Li, renormalising in the process energies, single-particle spectroscopic amplitudes  and wavefunctions (form factors) of virtual
	and resonant states.
	The fact that the cross section associated with the $2^-$ state $(s_{1/2} \otimes p_{3/2}(\pi))_{2^-})$ dominates over the 
	$1^-$ one ($s_{1/2} \otimes p_{3/2}(\pi))_{1^-})$ is because in the first case the coupling does not imply spin flip, while 
	it does so in the second one. The situation is naturally reversed concerning the  doublet 
	$(p_{1/2} \otimes p_{3/2}(\pi))_{1^+,2^+}$).
	
Similar calculations to the ones discussed above were carried out  but this time for the reaction $^9$Li(d,p) investigated at 2.36 MeV/A at the REX-ISOLDE 
facility \cite{Jeppesen:06}, making use of the optical potentials of this reference (see also \cite{Orrigo:09}). The results  provide an overall quantitative  description  of the 
experimental findings (Fig. 4).

	 Summing up, the  results shown in  Figs. \ref{fig2} and \ref{fig3}   dissipate the  possible doubts concerning the presence of a virtual $s_{1/2}$ state in the low--energy continuum spectrum of $^{10}$Li, and confirm the soundness of the picture at the basis of the description of $^{11}$Li provided in \cite{Barranco:01} and \cite{Potel:10} (see also \cite{Tanihata:08}). Within this scenario, Fig. 3c) and d)
constitute absolute strength function and differential cross sections predictions  with an estimated error of 10\%.

	\textit{Conclusions}. Structure and reactions, in particular when referred to a specific elementary mode of excitation (e.g. single--particle motion) and its specific probe (one--nucleon transfer), are two aspects of the same physics. Renormalized energies and wavefunctions (effective $Q$--values, spectroscopic amplitudes and formfactors) are the ``observables'', the meeting point between theory and experiment being absolute differential cross sections.
	
	For normal nuclei, structure essentially refers to bound states, reactions to continuum asymptotic waves. A distinction which becomes blurred in the case of exotic light bound halo nuclei like $^{11}$Be and $^{11}$Li. Just think in the $5/2^+$ resonance  in the first case (centroid $E_x=1.28$ MeV, width $\Gamma=100$ keV) and in the soft--dipole mode  in the second ($E_x\lesssim1$ MeV, $\Gamma=0.5$ MeV) , let alone on the virtual and resonant states in the case of $^{10}$Li and of its specific probe, $^9$Li(d,p)$^{10}$Li. 
	
	Making use of renormalized nuclear field theory of structure and reactions we find it similarly possible (trying) to provide a ``complete'' description of the structure and reaction process associated with $^{11}$Be and $^{11}$Li than with $^{10}$Li, in which case one is referring exclusively to continuum spectroscopy (structure) and reactions. The parameters used to calculate the bare single--particle levels of $^{10}$Li  were obtained by extrapolating those determined following the protocol presented in \cite{Barranco:17} 
in connection with the calculation of the $^{11}$Be spectrum applied also to the isotones $^{13}_6$C$_7$ and $^{12}_5$B$_7$ which display Mayer--Jensen sequence. 
The apparent puzzle--(hieroglyphic--) like position (see \cite{Cavallaro:17} and refs [7,8,10,11] therein, i.e.  \cite{Thoennessen:99,Zinser:95,Chartier:01,Simon:07} of the present Letter) of the continuous structure and reaction aspects of $^{10}$Li within the $N=6$, parity inverted scenario becomes readily understandable as the consequence of the choice of a restricted angular window associated with
low linear momentum transfer.
	
	Within the unified view adopted in the present letter, $^{10}$Li, $^{11}$Be and $^{11}$Li can be viewed as the top, middle and bottom texts of a rosetta--like stone dealing with parity inverted halo nuclei poised to acquire a permanent dipole moment. The uniqueness of the apparently different phenomena (``texts'') is due to the fact that they all emerge from the same underlying physics, namely: a) a quantal phase transition \cite{Sachdev:01} close to the crossing point (parity inversion) ; b) spontaneous symmetry breaking phenomenon (dipole instability) \cite{Anderson:84b}, and of their interplay. Another example of the relation existing between physical correct collective variables, emergent properties, transferability and effective lower dimensionality of many--body systems (\cite{Buchanan:15,Transtrum:15,Nikisic:16} and refs. therein). 

	Discussions with H. Lenske are gratefully acknowledged by RAB.
	F.B and E.V. acknowledge  funding from
the European Union Horizon 2020 research and innovation
program under Grant Agreement No. 654002.
F.B. thanks  the Spanish Ministerio de  Econom\'\i a y Competitividad and FEDER funds under project FIS2017-88410-P.
	
%
%
	

\bibliographystyle{unsrtnat}
\bibliography{nuclear_bib}
 \end{document}


\centerline{\bf Supplemental material}

{\it The bare  mean field potential}

The parameters of the bare mean field adopted in the various nuclei
are listed  in Table \ref{bare_para}. 
They  have been fitted in the even nuclei,
minimising the difference between the experimental energies of the $1/2^+,1/2^-$
and $5/2^+$ states  with the theoretical results obtained including the many-body
renormalisation effects.
In the case of the odd-odd nucleus $^{12}$B we have fitted the centroids 
of the multiplets (see below). In the case of  $^{10}$Li, the fit has been obtained
extrapolating the potentials obtained for the other nuclei and then adjusting the parameters
so as to reproduce the shape of the  strength function obtained in the transfer reaction
$^9$Li(d,p)$^{10}$Li by Cavallaro \textit{et al.} \cite{Cavallaro:17}, as discussed in the main text.  
We note that  in individual nuclei, fits of similar quality can be obtained with   
different families of parameters. We have privileged solutions having relatively large radii
and diffusivities, and evolving smoothly from one isotope to the next.

\begin{table}[h!]
\begin{center}
\begin{adjustbox}{max width=\textwidth}
\begin{tabular}{|c|c|c| c |  c | c |  c |  }
\hline 
&$\hbar \omega_{2^+}$ (MeV) & $\beta^n_2$ &$ V_{WS}$ (MeV) & $V_{ls} $ (MeV) 
& $a_{WS} $ (fm) &  $R_{WS}$  (fm)  \\ \hline
$^{10}$Li &3.37& 0.68 & 64 & 14 & 0.75& 2.10 \\ \hline
$^{11}$Be&3.37& 0.71 & 72   & 18  & 0.72 & 2.14 \\ \hline
$^{12}$B& 3.80 &  0.57 & 77&  22& 0.78  &2.18 \\ \hline
 $^ {13}$C & 4.4 & 0.46  & 82&  27 & 0.73 & 2.23     \\ \hline
 \end{tabular}
\end{adjustbox}
\end{center}
\caption{Parameters of the bare potentials adopted in the various isotopes considered, parameterized
according to Eq. (2.182) of ref. \cite{Bohr:69}.}
\label{bare_para}
\end{table}

\clearpage

{\it Deformation parameters}

\begin{figure}[h!]
\centerline{\includegraphics*[width=.8\textwidth,angle=0]{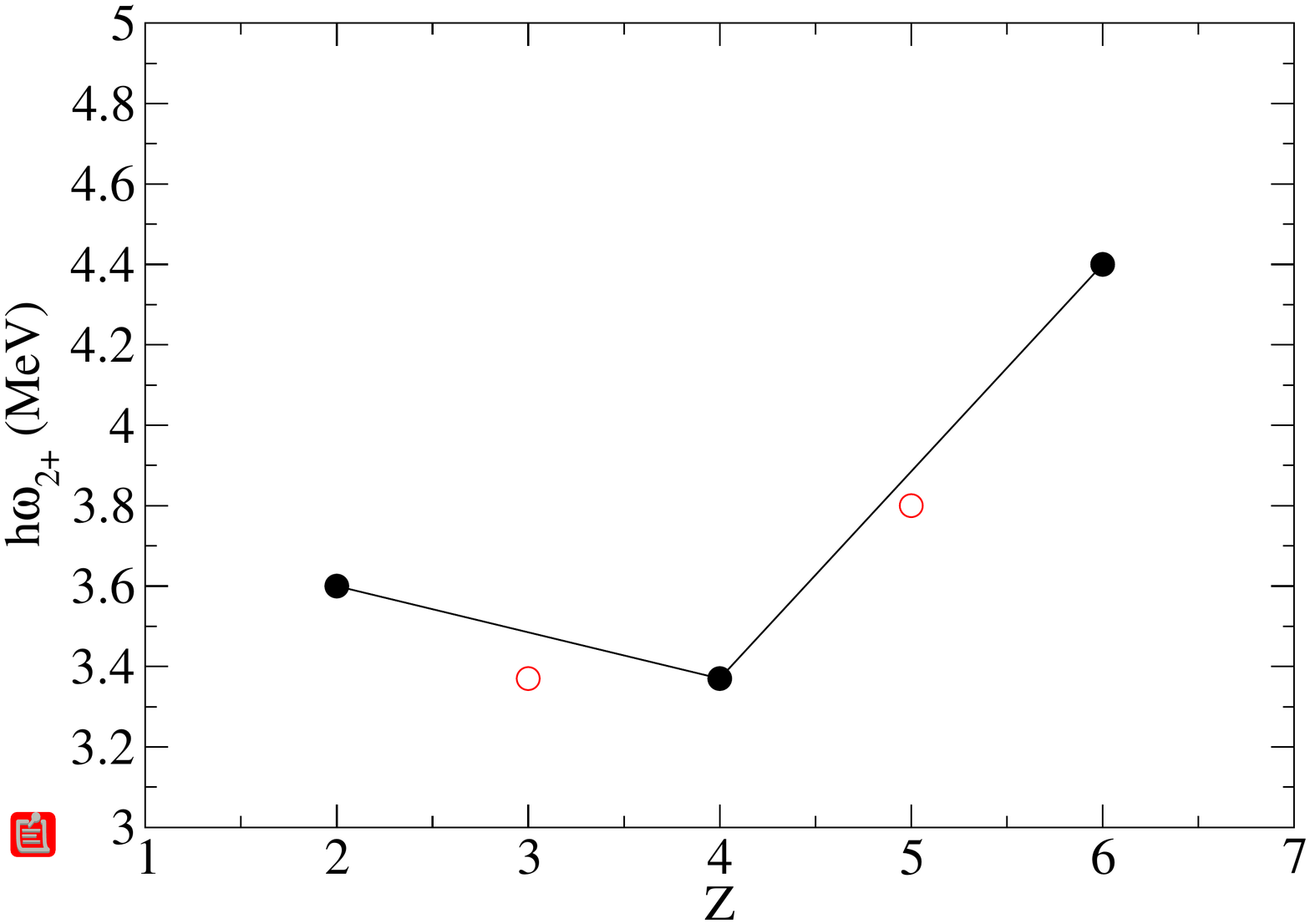}}
\caption{The experimental values of the $2^+$ energy  for the even-even  N=6 isotones $^8$He,$^{10}$Be and $^{12}$C  
are shown  as a function of the proton number (filled circles); 
the same values have been adopted in the calculations.  The values adopted 
in the calculations of  the odd-odd N=7 isotones $^{10}$Li and $^{12}$B   are shown by  open circles.  }
\label{ene_2+}
\end{figure}

\begin{figure}[h!]
\centerline{\includegraphics*[width=.8\textwidth,angle=0]{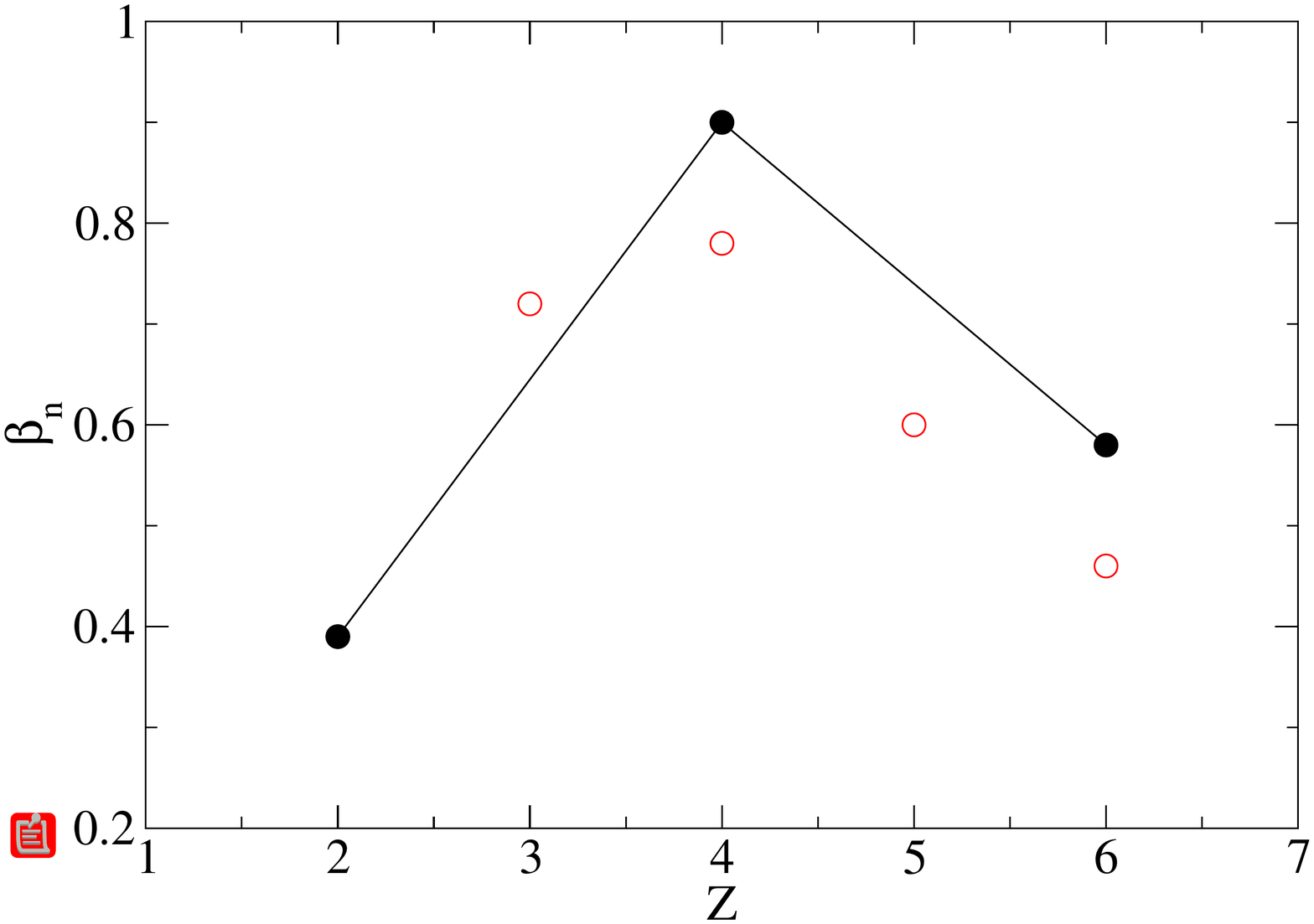}}
\caption{Experimental values of the neutron deformation parameter for the even-even N=6 isotones  $^8$He,$^{10}$Be and $^{12}$C (filled circles)  and values adopted 
in the calculations of the  N=7 isotones $^{10}$Li,$^{11}$Be, $^{12}$B and $^{13}$C   (open circles) as a function of the proton number. } 
\label{beta_2+}
\end{figure}

The renormalization of the neutron  single-particle states due to the coupling 
  to the low-lying collective quadrupole vibration is an essential ingredient of 
 our calculations of the structure of the $N=7$  isotones $_3^{10}$Li, $_4^{11}$Be,$_5^{12}$B,$_6^{13}$C.
 The properties of such dressing effects are determined by the energy $\hbar \omega_{2^+}$ and by the deformation parameter 
 $\beta_n$. The experimental data associated with the  $N=6$ even-even cores $_2^8$He, $_4^{10}$Be,$_6^{12}$C
 are  indicated  in Fig.   \ref{ene_2+}  ($\hbar \omega_{2^+}$ ) and Fig. \ref{beta_2+}   ($\beta_n$)   by filled circles.
In the case of $_4^{10}$Be,$_6^{12}$C   the value of $\beta_n$ is inferred from  the value of the 
electromagnetic $B(E2)$ and of the deformation parameter  obtained by inelastic (p,p') excitation.
In the case of $^8$He, the value of $B(E2)$ is not known and we have derived it from the values of $M_n/M_p$ and
$M_n$ reported in ref. \cite{Lapoux}. 

In our calculations the values of the neutron deformation parameter have been  fitted in each  calculated N=7 isotone
so as to optimize the properties of  the calculated energy spectrum as compared to experiments; they are shown 
in  Fig.  \ref{beta_2+} by empty circles. It is reassuring that the  values adopted
for  $_4^{11}$Be, and $_6^{13}$C lie close to the experimental values of the corresponding $_4^{10}$Be,$_6^{12}$C  cores,
and that the values adopted for the odd-odd nuclei $_3^{10}$Li and $_5^{12}$B interpolate smoothly between  them.
In our calculation of $_4^{11}$Be and $_6^{13}$C we have used the experimental values of the $2^+$ energy in 
$_4^{10}$Be and $_6^{12}$C. The values of  $\hbar \omega_{2^+}$ in the  odd-odd isotones  $_3^{10}$Li and $_5^{12}$B
have instead been fitted. Experimental and theoretical values are reported in Figs. \ref{ene_2+} and \ref{beta_2+}.

\clearpage

{\it Treatment of the neutron-proton interaction.}

This section presents a simple but quantitative  treatment of the effects related to the presence of the odd proton in $^{12}_5$B,
that can be directly extrapolated   to case  of $^{10}_3$Li.

The lowest states observed in the energy spectrum of  the odd-odd nucleus $^{12}$B are two doublets with angular momentum and parity ($1^+,2^+$) and
($1^-,2^-$) which can be naturally interpreted as arising from the coupling of the odd proton  lying on  the $1p_{3/2}$ orbital with
the odd neutron lying on the $1p_{1/2}$  ($\pi_{1p_{3/2}} \otimes \nu_{1p_{1/2}}$ ) or on the $2s_{1/2}$ ($\pi_{1p_{3/2}} \otimes \nu_{2s_{1/2}}$ ) weakly bound orbitals. 




The  doublets display a similar energy splitting equal to  approximately 0.95 MeV. The $1^+$ and the $2^-$ states  lie lowest in the two multiplets.
The relative position of the centroids  of the  $(1^+, 2^+)$ ($E_{cent}=0.59$ MeV)  and $(1^-, 2^-)$ doublets ($E_{cent}=2.02$)
 in $^{12}$B indicates a difference in the position of the $1p_{1/2}$ or on the $2s_{1/2}$  neutron states equal to $\Delta_{ps}$ = -1.43 MeV, to be compared with 
 with the value $\Delta_{ps}$ = -3.09 MeV  associated with the relative position of the states  $1/2^-$  (g.s.) and $1/2^+$ (3.09 MeV) in  $^{13}$C.

To calculate these splittings use has been made  of the schematic  two-body interaction 
$V_{np}= F_0 \delta(\vec r_1- \vec r_2) (1-\alpha+\alpha \sigma_1 \cdot \sigma_2)$,
originally introduced  in order to interpret  Nordheim's rule  for the spin of the ground state of odd-odd nuclei
\cite{Heyde,Deshalit,Schwartz}. 
The magnitude of the splittings within each $\pi(p_{3/2}) \otimes \nu(nl_j)$ multiplet is determined in first order by the diagonal matrix element of $V_{np}$ 
in the neutron-proton wavefunction, and  the relative ordering of the levels  is determined by the parameter $\alpha$. 
For  values of the parameter $\alpha \approx 0.5$, one reproduces the experimental  ordering of the  states in $^{12}$B
within the two doublets (the states $1^+$ and  2$^-$ are the lowest ones).  

The action of the two-body interaction we have adopted is in general different, according to whether the neutron and 
proton levels involved are  of hole- or of  particle-character. In  the case of one-hole and one-particle states, the shift caused by the spin-independent part changes sign, while 
 the $\sigma_1 \cdot \sigma_2$ part does not change. 
 However, the ordering within the  $\pi(p_{3/2}) \times \nu(2s_{1/2}) $ and $\pi(1p_{3/2}) \times \nu(1p_{1/2}) $  doublets   in $^{10}$Li is predicted to be the same as 
 within the $\pi(1p^{-1}_{3/2}) \times \nu(2s_{1/2}) $  and $\pi(1p^{-1}_{3/2}) \times \nu(1p_{1/2}) $ doublets   in $^{12}$B.
 In fact ,the $1p_{1/2}$ and $2s_{1/2}$ states 
can be considered as particles or as holes 
(one particle in a two-level orbital), and for symmetry the contributions of the spin-independent 
part does not produce  any splitting, but only a level shift which is taken into account by the mean field.
As a consequence, the splittings caused by these interaction are expected to be the same, apart from the 
difference arising from the different radial form factors.
 This is a very peculiar characteristic of the multiplets based on  $s_{1/2}$ or $p_{1/2}$ orbitals. 
In general the situation is different , as for example in the  case of the multiplets based on the $1d_{5/2}$  orbital, 
which are predicted to  show inverted splittings in $^{10}$Li and $^{12}$B.

For a more quantitative calculation,  the interaction should  be diagonalized in a space of good total angular momentum $J$ 
generated by all the possible combinations of the odd proton and the odd neutron orbitals that may couple to $J$.
We will make two simplifying assumptions.
Our first approximation will be to freeze the proton in the $1p_{3/2}$ orbital, which is justified due to  its large binding energy.
Our second approximation will be to neglect the mixing  between different multiplets, an approximation which is justified 
by their relatively large energy separation. 
 In particular, we will not mix the $2^-$ state produced  by coupling the proton with  the neutron moving in  the $2s_{1/2}$ orbital, 
 with the $2^-$ states obtained  from the coupling with the neutron moving in the $1d_{5/2}$ or in the $1d_{3/2}$ orbitals. 
 With these two approximations, the problem becomes  equivalent to a $ J(lj)$-dependent one-body problem.
The neutron wavefunction  may  be obtained solving the  Schr\"odinger equation adding a $J(lj)$-dependent potential given by
$\Delta V_{J(lj)} = F_0 |\phi_{p_{3/2}}(r)|^2 < (1-\alpha+\alpha \sigma_1 \cdot \sigma_2)>_{J(lj)}$  to the bare potential used to fit the centroid, making use of
the radial proton wave function $\phi_{p_{3/2}}(r)$.

For simplicity we have substituted the radial form factor  $\phi_{p_{3/2}}(r)$ by  a  Wood Saxon 
potential having the same geometry as the bare potential used to fit the centroid,
determining its depth so as to  reproduce the experimental value of the position of the $1^+$ level,  leading 
to $\Delta V_{1(s_{1/2})}=$ -2.6 MeV. The depth of the potentials used for the other $J(lj)$ states   have been scaled according to the value of 
  $< (1-\alpha+\alpha \sigma_1 \cdot \sigma_2)>_{J(lj)}$. 
We have adopted the same procedure in the case of $^{10}$Li, but starting from the bare  WS potential with parameters  extrapolated from those determined 
in $^{13}$C and $^{11}$Be. The  comparison with the data of the reaction discussed in the main text 
favours somewhat smaller values of  $\Delta V_{1(s_{1/2})}$. We have adopted $\Delta V_{1(s_{1/2})}=$ -1.8 MeV which leads to a rather good reproduction of
the data both in in $^{12}$Be and $^{10}$Li.

 